\documentclass[sigconf,nonacm]{acmart}

\renewcommand\footnotetextcopyrightpermission[1]{} 
\AtBeginDocument{%
  \providecommand\BibTeX{{%
    \normalfont B\kern-0.5em{\scshape i\kern-0.25em b}\kern-0.8em\TeX}}}

\setcopyright{none}
\copyrightyear{2019}
\acmYear{2019}
\acmDOI{10.1145/1122445.1122456}

\acmJournal{IMWUT}
\acmVolume{37}
\acmNumber{4}
\acmArticle{111}
\acmMonth{8}

\usepackage{tabularx}
\usepackage{booktabs}
\usepackage{hyperref}
\usepackage{graphics}
\usepackage{graphicx}
\usepackage{color,soul}
\usepackage{algorithm}
\usepackage[noend]{algpseudocode}
\usepackage{xcolor}   
\usepackage{pifont}   
\usepackage{subcaption}  
\usepackage{multirow}
\usepackage{wasysym} 
\usepackage{color, colortbl}
\usepackage{balance}
\usepackage{verbatim}
\usepackage{xspace}
\usepackage{balance}
\usepackage{amsmath, amsthm, amssymb, amsfonts}

\newcommand{\pioti}[1]{IoT Inspector}

\newcommand{\TableHeader}[1]{\textit{\small #1}}

\newcommand{\Section}[1]{Section~\ref{sec:#1}}
\newcommand{\Figure}[1]{\textbf{Figure~\ref{fig:#1}}}
\newcommand{\Table}[1]{\textbf{Table~\ref{tab:#1}}}
\renewcommand{\paragraph}[1]{\vspace{5pt}\noindent\textbf{#1: }}
\newcommand{\paragraphnocolon}[1]{\vspace{5pt}\noindent\textbf{#1 }}

\newenvironment{prettylist}{
  \raggedright
  \begin{list}{
    \footnotesize\raisebox{0.1mm}{\small\ding{118}}
  }{
    \setlength\topsep{1ex}
    \setlength\leftmargin{22pt}
    \setlength\rightmargin{0pt}
    \setlength\itemsep{2pt}
    \setlength\parskip{0pt}
    \setlength\parsep{0pt}
    \setlength\itemindent{0pt}
  }
}{
  \end{list}
}

\begin{document}

\title{IoT Inspector: Crowdsourcing Labeled Network Traffic\\from Smart Home Devices at Scale}

\author{Danny Yuxing Huang$^1$ \:\: Noah Apthorpe$^1$ \:\: Gunes Acar$^2$ \:\: Frank Li$^3$ \:\: Nick Feamster$^4$}
\affiliation{$^1$ Princeton University \:\: $^2$ KU Leuven \:\: $^3$ Georgia Institute of Technology \:\: $^4$ University of Chicago}
\authorsaddresses{}

\renewcommand{\shortauthors}{}

\begin{abstract}

The proliferation of smart home devices has created new opportunities for
    empirical research in ubiquitous computing, ranging from security and
    privacy to personal health. Yet, data from smart home deployments are hard
    to come by, and existing empirical studies of smart home devices typically
    involve only a small number of devices in lab settings. To contribute to
    data-driven smart home research, we crowdsource the largest known dataset
    of labeled network traffic from smart home devices from within real-world
    home networks. To do so, we developed and released \pioti{}, an
    open-source tool that allows users to observe the traffic from smart home
    devices on their own home networks.  Since April 2019, 4,322 users have
    installed \pioti{}, allowing us to collect {\em labeled} network traffic
    from 44,956 smart home devices across 13 categories and 53 vendors. We
    demonstrate how this data enables new research into smart homes through
    two case studies focused on security and privacy. First, we find
    that many device vendors
    use outdated TLS versions and advertise weak
    ciphers. Second, we discover about 350 distinct third-party
    advertiser and tracking domains on smart TVs.  We also highlight other
    research areas, such as network management and healthcare, that can take
    advantage of \pioti{}'s dataset.  To facilitate future reproducible
    research in smart homes, we will release the \pioti{} data to the public.

\end{abstract}



\settopmatter{printfolios=true}

\maketitle

\section{Introduction}
\label{sec:introduction}

Internet-connected consumer devices, also known as smart home or
Internet of things (IoT) devices, have seen increasingly widespread adoption
in recent years. These new technologies create new challenges and research
opportunities for ubiquitous computing. Conventional challenges include
security (e.g., distributed denial-of-service attacks by IoT
botnets~\cite{usenix_mirai}); privacy (e.g., toys transmitting sensitive
information about children to third parties~\cite{chu2019security}); and
device inventory and management (e.g., determining what devices are connected
to a network~\cite{DeviceMien}). Ultimately, data about smart home
devices---and the usage of these devices---holds tremendous opportunities
for understanding how people use IoT technologies and for designing new
ubiquitous computing applications that rely on the collection or analysis of
data from these devices.

However, this research needs large amounts of labeled data from
smart home devices, which is challenging to obtain at scale for several reasons:

\textbf{(i) Scaling challenges.} According to one estimate~\cite{gartner_iot}, there
are more than 8 billion Internet-connected devices in the world. Many of these
devices are on private home networks~\cite{usenix_zakir_iot}. Yet, analysis of
smart home devices often requires either physical or local network access to
the devices themselves; as a result, much of the existing work operates in
small-scale lab environments~\cite{apthorpe2018keeping, wood2017cleartext}.
Researchers have occasionally deployed custom hardware in consumer homes to
gather information about devices in homes~\cite{bismark,schmitt2018enhancing},
but these types of deployments often require significant effort, since they
require users to install a (sometimes costly) physical device on their home
networks.  Another approach is to scan the Internet for exposed
devices~\cite{usenix_mirai}.  However, this approach omits devices behind
gateway routers that act as network address translators (NATs).

\textbf{(ii) Labeling challenges.} Absent a large corpus of
ground-truth device labels, researchers often can infer the identities of only
a limited set of devices~\cite{feng2018acquisitional}.  Researchers have
previously published analyses of proprietary data from Internet-wide scans have been
analyzed~\cite{usenix_zakir_iot}, but these datasets are not public and
typically do not have specific or reliable device labels.

These limitations make it difficult to carry out empirical ubiquitous
computing research based on data from real smart homes, ranging from
measurements of security/privacy violations in the
wild~\cite{apthorpe2018keeping, wood2017cleartext} to training machine
learning algorithms for modeling device
behaviors~\cite{mirsky2018kitsune,IoT_Sentinel} or inferring device
identities~\cite{DeviceMien,ProfilIoT}.

\textbf{Our solution.}
We have constructed the largest known dataset of labeled
smart home network traffic by developing and releasing an
open-source application, \textit{\pioti{}}, that crowdsources the data from within home networks in a
way that respects user privacy (\Section{implementation}).\footnote{\url{https://iot-inspector.princeton.edu}} Our Institutional
Review Board (IRB) has approved the study. Since we released the software on
April 10, 2019, \pioti{} has collected network traffic from 4,322
global users and 44,956 devices, 12,690 of which have user-provided labels.
We have validated the correctness of these labels
against external information; we discuss the challenges of label validation
and our validation approach  in
\Section{labels}.
The data are also still growing, as users are actively downloading and using
\pioti{} at the time of writing.

This unique dataset will enable many types of research that
have otherwise suffered from limited scale and labels.
Similar to how ImageNet~\cite{deng2009imagenet} advanced the field of
computer vision, we hope to contribute to smart home research
by providing our data to expand the
scope of empirical analyses and develop more generalizable or realistic
models.
Since we released \pioti{},
seven research groups have contacted us to ask about using the data in a wide
variety of ways, including:
\begin{prettylist}
    \item Training machine learning models for device identification and anomaly detection. 
    \item Measuring the security risks of IoT messaging systems such as MQTT~\cite{mqtt}. 
    \item Identifying third-party services in the smart home ecosystem. 
    \item Inferring and labeling human activities to understand the privacy risks of devices.
    \item Facilitating the collection of data concerning the habits of
        individuals in homes (e.g., eating, sleeping, screen time) to help
        answer questions related to personal health. 
\end{prettylist}

To demonstrate the potential of the \pioti{} dataset for ubiquitous computing
research, we analyzed a subset of the data collected between April 10
and May 5, 2019 to study the pervasiveness of two security \& privacy issues
plaguing smart homes: incorrect encryption of
network traffic and communication with third-party
advertisers and trackers.
During this period, 25\% of \pioti{} users collected
at least 2.8 hours of traffic each, and 10\% of these active users collected
at least 12.4 hours of traffic each. Additionally, 1,501 users manually
labeled the identities of 8,131 devices across 53 manufacturers
(\Section{overview}).

We found that although 46 out of 53
observed vendors use TLS on their devices, devices from several popular
vendors use outdated TLS versions or advertise
insecure ciphers (\Section{encryption}).
We also found that 14 out of 19 observed smart TV vendors communicate third-party
advertising and tracking services, identifying 350 distinct third-party advertiser and tracking domains.
This is the first known identification of specific third-party
services that could collect and aggregate behavioral and lifestyle data across
a range of smart devices that a user might have in his or her home (\Section{data-aggr}).

In addition to these
two case studies, we discuss other types of empirical ubiquitous
computing research that
the \pioti{} dataset enables, ranging from security and privacy to network management and
healthcare (\Section{future}).

This work makes the following contributions:
\begin{prettylist}

    \item We have crowdsourced the largest known dataset of labeled,
        real-world, smart home network
        traffic and device labels using \pioti{}, an open-source tool that we
        developed to help gather this data at scale..

    \item Through an initial analysis of the dataset, we discovered widespread security
        and privacy with smart home devices, including insecure TLS implementation and pervasive use
        of tracking and advertising services. Such information is uniquely
        available in the \pioti{} dataset.
        In addition to these preliminary case studies, we highlight other
        types of ubiquitous computing research that can use the dataset.

    \item \textbf{We make the dataset available to interested researchers} (Section~\ref{sec:data-release}).
    	This includes the anonymized network traffic (in the form of
        $\langle$device identifier, timestamp, remote IP or hostname, remote
        port, protocol$\rangle$) and device labels (in the form of $\langle$device identifier,
        category, vendor$\rangle$).\footnote{\url{https://iot-inspector.princeton.edu/blog/post/faq/\#for-academic-researchers}}
\end{prettylist}

\section{Related Work}
\label{sec:related}

We first discuss existing techniques to obtain large, labeled traffic
datasets and their relation to \pioti{} (\Section{related:data}). We then
describe previous and ongoing smart home studies that could benefit
from a large-scale, labeled dataset such as the one \pioti{} has collected
(\Section{related:existing}).

\subsection{Crowdsourcing labeled traffic datasets at scale}
\label{sec:related:data}

Existing techniques to obtain labeled network traffic at scale face multiple
challenges. In particular, lab studies are restricted to a small set of
devices~\cite{apthorpe2018keeping, wood2017cleartext}, while Internet-scanning
omits devices behind NATs and often produces limited device
labels~\cite{usenix_mirai,feng2018acquisitional}.

\paragraph{Hardware-based approaches}
We design \pioti{} to crowdsource the network traffic and labels
of smart home devices from a large user population, following in the footsteps of a number
of previous crowdsourcing studies. For example, multiple researchers have deployed custom
routers to collect the participants' home traffic: Chetty et
al.~\cite{why_slow} developed Kermit, a router-based tool, to help users
diagnose slow network connectivity. BISmark~\cite{bismark, behind_nat}
collected network performance characteristics through deploying routers in
home networks. NetMicroscope~\cite{schmitt2018enhancing} analyzed the quality
of video streaming services through custom routers in participants' home
networks. However, unlike \pioti{}, the hardware-based approaches used in these
studies are difficult to scale to more users
due to the cost of hardware and shipment.

\paragraph{Software-based approaches} We are not aware of any
software tools other than \pioti{} that collect smart home traffic at scale.
Netalyzr~\cite{netalyzr} was a web-based application that helped users analyze
home network performance and also gathered network statistics from
99,000 different public IP addresses. DiCioccio et al.~\cite{probe_pray}
developed HomeNet Profiler~\cite{homenet_profiler} to explore how effectively
UPnP could be used to measure home networks. While software tools are
typically easier to deploy than hardware routers, most such tools
have actively probed the home network (e.g., by performing a ``scan'') rather
than passively collecting traffic. 

\pioti{} combines the benefits of hardware and software data collection
platforms. By designing \pioti{} as a software tool, we avoid some of the
deployment barriers that router-based studies face. We also develop \pioti{}
to behave like a router and intercept network traffic via ARP spoofing
(\Section{arp_spoofing}), thereby building a dataset of smart home network
traffic at scale. Furthermore, we draw inspiration from
Netalyzr~\cite{netalyzr} and design \pioti{} to benefit users, with the goal
of promoting participation and user engagement (\Section{benefiting_users}).
At the same time, we make user privacy our first-order concern
(\Section{privacy}) much as in previous work~\cite{bismark}.

\subsection{Smart home research that could benefit from \pioti{} data}
\label{sec:related:existing}

The increasing prevalence of smart home devices has spurred researchers to
investigate a variety of problems using empirical methods. These studies have
typically relied on either small-scale laboratory-based data, or proprietary
datasets.

\paragraph{Discovering security and privacy violations} Past work has explored
security and privacy issues of a small set of smart home devices in lab
settings. Chu et al. and Sasha et al.~\cite{chu2019security,
shasha2018playing} found a variety of security flaws in IoT children's toys;
Wood et al.~\cite{wood2017cleartext} found cleartext health information in
home IoT medical device communications; and Acar et al.~\cite{iot_web_attacks}
presented web-based attacks on smart home devices that host local webservers,
demonstrating their real-world applications on seven home IoT devices (e.g.,
Google Home and Chromecast). A larger dataset of labeled network traffic 
enable the study the problems across a much wider array of devices and
vendors.

Other studies have relied on data from actively ``scanning'' devices on the
Internet or in home networks.
Antonakakis et al.~\cite{usenix_mirai} scanned the
Internet and identified public-facing devices compromised by the Mirai botnet;
Kumar et al.~\cite{usenix_zakir_iot} used a proprietary dataset from an
antivirus company to discover vulnerable devices within home networks. Despite
the scale of these studies, researchers do not have reliable labels of
device types and vendors; rather, they could only \textit{infer} the device
identities using a variety of signals (e.g., based on HTTP response
headers~\cite{feng2018acquisitional}, default passwords~\cite{usenix_mirai},
or proprietary rules~\cite{usenix_zakir_iot}). In contrast, lab studies
permit knowledge of device types and vendors but are limited to much smaller
scale.  \pioti{} allows the collection of a large, labeled dataset of network
traffic from devices that are deployed in real networks.

\paragraph{Modeling device activities} Other past work has applied machine
learning identify unknown devices~\cite{DeviceMien,ProfilIoT} and detect
anomalous behaviors~\cite{mirsky2018kitsune,IoT_Sentinel}. These studies used
a small number of devices in lab settings for training and validation. It is
unclear, however, whether the models would be equally effective if tested in
real-world settings, with a larger set of devices as being used by real humans.

\section{Crowdsourcing Smart Home Network Traffic at Scale}
\label{sec:implementation}

In this section, we describe the design and implementation of \pioti{}
to crowdsource labeled network data at scale.
We developed \pioti{}, an open-source tool that consumers can download on
their computers at home to analyze the network activities of their smart home
devices. To attract users to participate in this crowdsourcing effort, we
designed \pioti{} in a way that makes setup easy; our goal was to make the
application as close to ``one click'' as possible.  Users can run \pioti{} on
macOS- and Linux-based computers\footnote{The Windows version is still under
development at the time of writing.} without dedicated hardware such as custom
routers. Furthermore, we designed \pioti{} to promote user engagement by
showing a real-time analysis of their smart home network traffic on the user
interface, which allows users to identify potential security and privacy
problems. With user consent, \pioti{} anonymizes and uploads the network data
and device labels to our
server, where we preprocess the data for researchers to
analyze.

\subsection{Designing a software tool to capture traffic} \label{sec:arp_spoofing}

Many home network measurement platforms~\cite{bismark,schmitt2018enhancing}
require participants to first obtain custom routers to collect
network traffic. This requirement, however, limits the scale of such studies
due to the cost of hardware and shipment.

To minimize the setup cost and facilitate deployment at scale, we design
\pioti{} to be a software-based data collection tool that users can install in
a relatively small number of steps. First, the user needs a macOS- or
Linux-based computer that is connected to the smart home network. From
\pioti{}'s website, the user can download the precompiled executable or the
source code.\footnote{Users have to accept macOS's warning that the app
is not from the official AppStore. We cannot submit the app to the AppStore
because it does ARP spoofing.} The executable includes all necessary
platform-dependent libraries, so that the user can launch \pioti{} without
needing to install additional packages. When a user runs \pioti{} for the first time,
it displays a consent form---approved by our university's IRB---that details
what data \pioti{} collects and that \pioti{} poses no more than
minimal risk to users.

Upon the user's consent, \pioti{} automatically discovers devices on the
network and captures traffic from select devices, as outlined below:

\begin{figure}[t] \centering
\fbox{\includegraphics[width=\columnwidth]{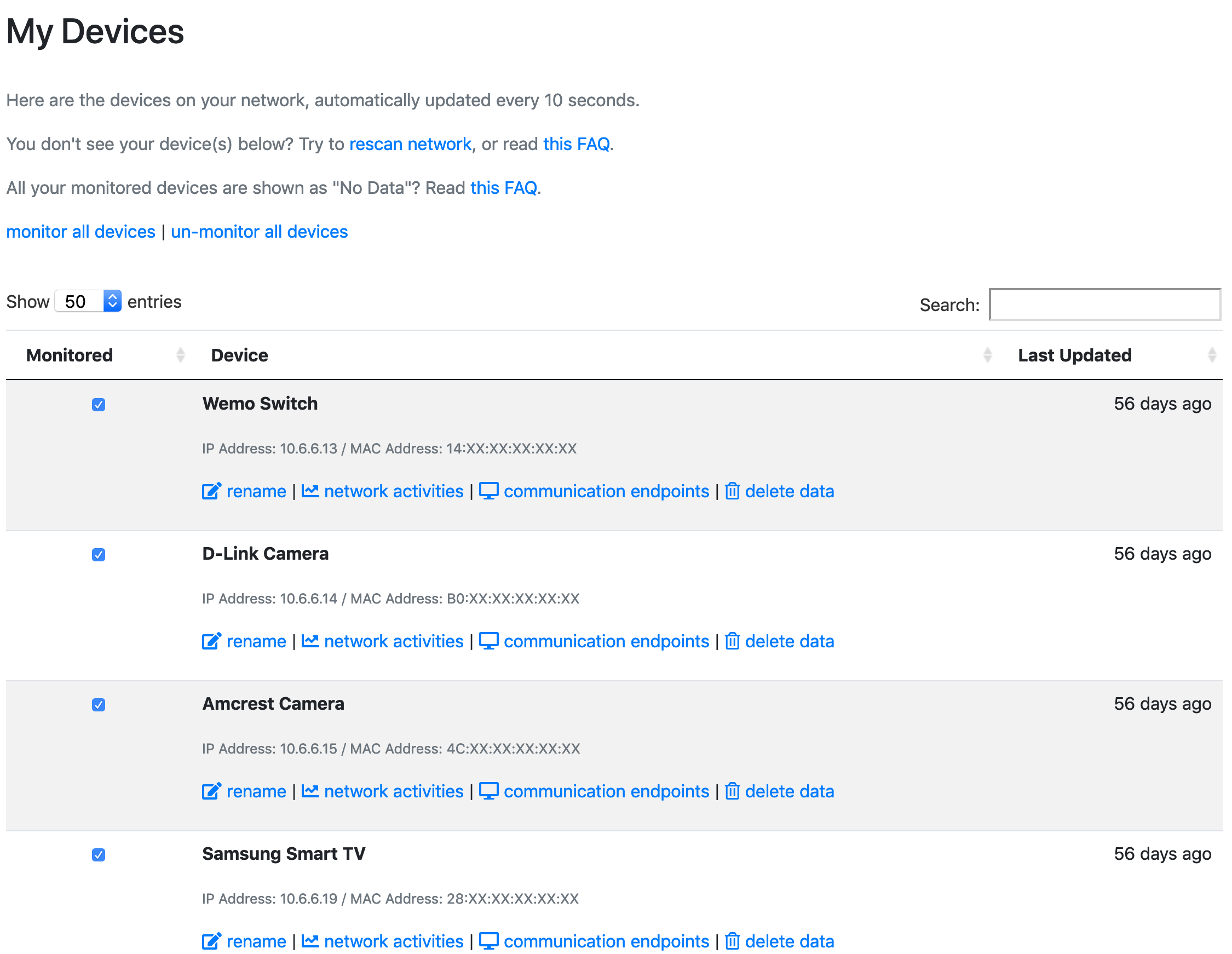}} \caption{A
screenshot of \pioti{}'s user interface that shows a list of devices on the
network. Using the checkboxes, users can select which devices to monitor, i.e., from which \pioti{} can capture network traffic} \label{fig:devices} \end{figure}

\paragraph{Discovering devices via ARP scanning} Upon launch, \pioti{}
automatically sends out ARP packets to all IP addresses in the local subnet to
discover devices. At the same time, \pioti{} opens the user interface (UI) in
a browser window that shows a list of device IP addresses and MAC addresses
currently on the network. We show an example screenshot of this UI in
\Figure{devices}. From this UI, users have to explicitly indicate which
of the listed devices \pioti{} is allowed to monitor (i.e.,
collect traffic). To help users choose what devices to monitor,
\pioti{} also displays the \textit{likely} identities of individual devices,
using external data sources such as the IEEE Organizationally Unique
Identifier (OUI) database (which shows the manufacturers of network
chips~\cite{ieee-oui}) along with mDNS and SSDP
announcements that may include a device's identity (as collected by Netdisco~\cite{netdisco}).

\paragraph{Capturing traffic via ARP spoofing} By default, \pioti{} only ARP scans
the network to discover devices. For \pioti{} to capture any device
traffic, a user would have to explicitly indicate which device(s) to monitor
from the device list (\Figure{devices}).

For each monitored device, \pioti{} continuously sends two ARP spoofing
packets every two seconds, similar to Debian's \texttt{arpspoof}
utility~\cite{arpspoof}: one packet is sent to the monitored device using the IP
address of the router as the source, and one packet is sent to the router using
the IP address of the monitored device as the source. In this way, \pioti{}
can intercept all traffic between the monitored device and the router.

The ARP spoofing packets are unlikely to consume significant bandwidth,
although network latency is likely to be increased due to packets taking extra hops
to go through \pioti{}. Each ARP packet is typically 42 bytes. If there are $N$
monitored devices (excluding the router) on the local network, then \pioti{}
would need to send out $2 (N + (N-1) + (N-2) + ... + 1) = (N+1)N$ packets
every two seconds, or $21(N+1)N$ bytes per second. In a home network of 50
devices (which is the upper limit for \pioti{} by default), for instance, the
bandwidth overhead would be 53.6 Kilobytes/second.

Through a combination of ARP scanning and spoofing, \pioti{} is able to
discover devices and capture their traffic in a way that requires minimal user
engagement and no dedicated hardware. Using this captured
traffic, we can generate a dataset for research
(\Section{collecting_traffic_and_labels}) and promoting user engagement
(\Section{benefiting_users}).

\subsection{Collecting network traffic and device labels}
\label{sec:collecting_traffic_and_labels}

\pioti{} collects two types of data: network traffic and device labels.

\paragraph{Network traffic}
Users choose which devices to \textit{monitor} (\Figure{devices}), such that
packets to and from the monitored devices are
captured by the computer that runs \pioti{}. \pioti{} parses the relevant
fields from the captured packets using the Scapy Python library, removes
sensitive information, and uploads the resulting data to the database server
at five-second intervals. Specifically, \pioti{} collects the following data:

\begin{prettylist} \item SHA-256 hashes of device MAC addresses, using a
    secret salt\footnote{\pioti{} does not share the secret salt with us.}
that \pioti{} randomly generates upon first run.
\item Manufacturer of the device's network chipset, based on the first 3
        octets of the MAC address (i.e., OUI).
    \item DNS requests and
        responses.
    \item Remote IP addresses and ports.
    \item Aggregated flow
        statistics, such as the number of bytes sent/received over
    five-second windows.
\item Data related to device identities, including
        SSDP/mDNS/UPnP messages, HTTP User-Agent strings, and hostnames from
        DHCP Request packets, that are useful for validating device identity
        labels entered by users (\Section{labels}).
    \item TLS Client Hello messages.
\item Timezone of the computer running \pioti{} \end{prettylist}.

\paragraph{Device labels} Recorded network traffic alone is typically insufficient
for research, as it is often necessary to characterize network activities that
are specific to certain models or types of devices.
\begin{figure}[t] \centering
    \fbox{\includegraphics[width=\columnwidth]{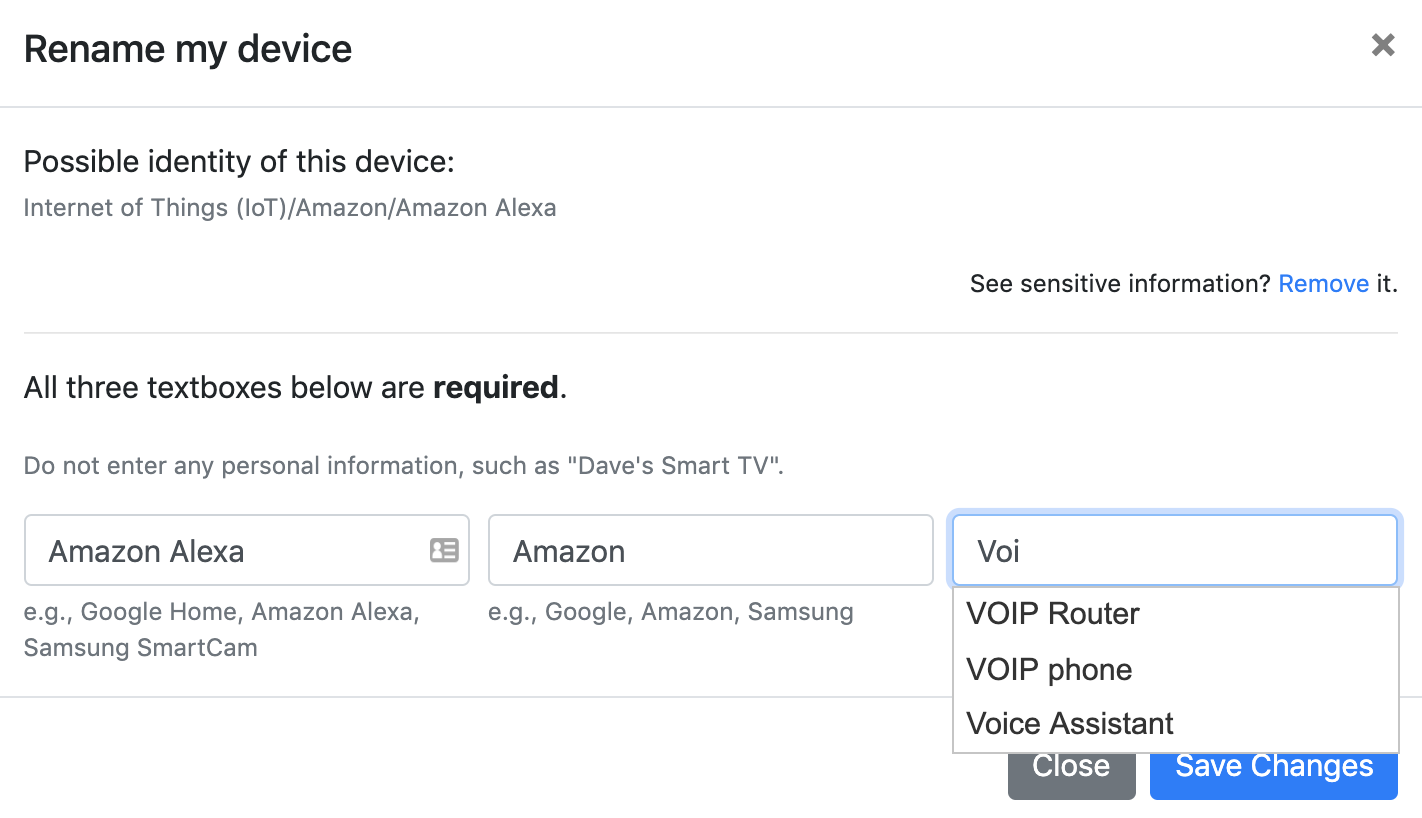}} \caption{A
    screenshot showing a user entering the identity information of a device
    into a dropdown textbox with auto-complete.} \label{fig:device_label}
\end{figure}
\pioti{} therefore asks users to voluntarily label their devices' identities.
From \pioti{}'s UI, users can enter the name (e.g., ``Roku TV''), category
(e.g., ``TV''), and vendor (e.g., ``Roku'') for one or more of their devices.
\pioti{} provides dropdown textboxes with auto-complete capabilities, so that
users can select from a list of known labels. If, on the other hand, the
desired labels are not in the dropdown lists, users can enter free text. We
show an example of the device labeling interface in \Figure{device_label}.

\pioti{} uploads the device labels along with the network traffic data to a
central database hosted at our institution. We use this dataset to
investigate two security and privacy issues of smart home devices within and
across device categories and vendors (\Section{findings}).

\subsection{Protecting privacy of others in household} \label{sec:privacy}

The design of \pioti{}, along with our data collection, storage, and retention
policies/practices, has been approved by our  university's IRB. We follow
industry-standard security and privacy practices. For example, each instance
of \pioti{} uploads the captured network data to our central server via HTTPS,
and we store this data on a secure, fully updated server hosted on our
institution's network. \pioti{} only collects the data outlined in
\Section{collecting_traffic_and_labels}.

Nonetheless, the data collected may inadvertently contain sensitive
information. As such, we designed \pioti{} to allow a user to retroactively
remove select data. For example, a device could be involved in sensitive
activities, or a user may have accidentally monitored a medical device; in this case,
the user can delete all data associated with this device from our server directly through
\pioti{}'s UI. Additionally, collected DHCP or SSDP messages may include a user's
identity (e.g., ``Joe's TV''); in this case, the user can have \pioti{} remove
all DHCP and/or SSDP messages from a specific device from our server.

Furthermore, \pioti{} may pose privacy risks to other people on the same
network who do not use \pioti{}. Although ARP spoofing makes
make it easy for \pioti{} to start capturing traffic, this design could also
potentially make it easy for a user to analyze sensitive activities of other
people on the same network.

To increase the barrier of such malicious activities, we design \pioti{} such
that it does not upload any traffic from devices that show signs of being
general-purpose computing devices, such as phones, tablets, or computers. We
make this determination based on two data sources: (i) the HTTP User Agent
string (which is often missing due to reduced adoption of unencrypted HTTP);
and (ii) the FingerBank API~\cite{fingerbank}, a proprietary service that
takes as input the first 3 octets of a device's MAC address, along with a
sample of five domains contacted, and outputs the device's likely identity. By
parsing this output and looking for specific keywords such as ``phone,''
``macOS'', ``Android,'' or ``Windows,'' we infer whether the device is potentially
a smart home device or a general purpose computer.

It is possible that \pioti{} may mistake an actual smart home device for a
computer (e.g., due to false positives of FingerBank's API). Users can
manually correct this mistake by following the instructions on the \pioti{} UI and
entering the device's MAC address (e.g., often printed on the device itself,
or displayed on the settings menu if the device has a screen), thereby
demonstrating that the user likely has physical access to the device. We
admit, however, that this design merely increases the barrier for a malicious
user; it does not completely prevent an advanced user from ARP-scanning the
network, obtaining the MAC address of a targeted device, and bypassing this
protection. At the same time, it is worth noting that a user who is advanced
enough to bypass \pioti{}'s protections is likely sophisticated to perform
network traffic capture and monitoring {\em without the help of \pioti{} in
the first place}.

\subsection{Keeping users engaged} \label{sec:benefiting_users}

Our goal is to not only make this dataset useful to researchers; it should
also provide users with insight on their smart home networks. This draws from
Netalyzr's experience that providing benefits to users also boosted user count
and engagement~\cite{netalyzr_experience}.

To this end, we set up an automated script on the server to preprocess the
data collected, produce tables and charts in real time, and push these
visualizations back to the front end for users to discover potential security and
privacy issues.

\begin{figure}[t] \centering
    \fbox{\includegraphics[width=\columnwidth]{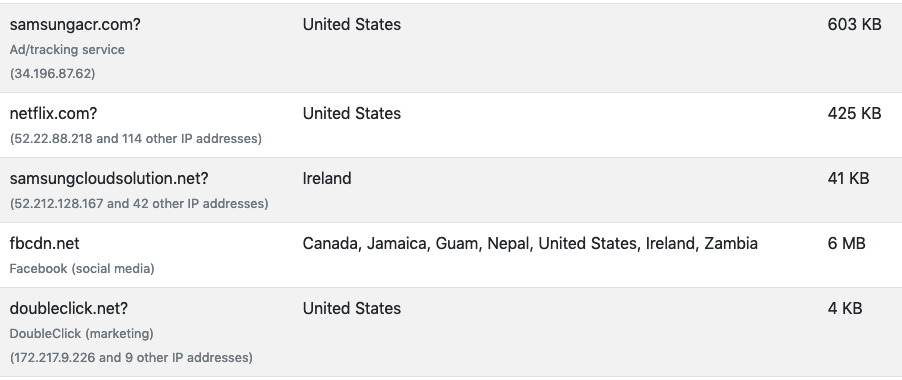}}
    \caption{A screenshot showing a table of endpoints, along with the
    countries and the number of bytes sent and received for each endpoint, for
    a Samsung Smart TV in our lab.} \label{fig:endpoint_table} \end{figure}

\paragraph{Preprocessing data} The data preprocessing pipeline involves two
steps: (1) aggregating the collected packets into flows (i.e., same source and
destination IP addresses, ports, and protocol) at 5-second
intervals; and (2) identifying the remote endpoints that communicate with the
monitored devices.

We identify remote endpoints because, by default, each packet collected only
shows the communication between a monitored device and some remote IP address.
An average user, however, may not know which real-world entity is associated
with the IP address. As such, our automated script attempts to first find the
hostname that corresponds to the remote IP address based on past DNS responses
or the SNI field from previous TLS Client Hello messages. It is possible that
one or both of DNS and SNI might be missing in the dataset; for example, a
user could have started running \pioti{} after the relevant DNS and SNI
packets were sent/received and thus \pioti{} would fail to capture these
packets. In the case where we do not observe DNS or SNI data for a particular
IP address, we infer the hostname based on passive DNS~\cite{farsight} or
reverse DNS (i.e., PTR records).

It is possible that the hostnames themselves may not always be indicative of
the endpoint's identity. For instance, an average user may not know that
\texttt{fbcdn.net} is a Facebook domain. To help users learn about the
identities of endpoints, we use the webXray Domain Owner List to turn hostnames into human-readable
company names~\cite{webxray}. We also use the Disconnect list to label
hostnames that are known to track users or serve
advertisements~\cite{disconnect}. We further complement this information with
a manually compiled database of common ports; for instance, if a device
contacts an IP address with  destination port 123, \pioti{} shows the user
that the remote service is likely an NTP time server. We show an example of
these human-readable labels in \Figure{endpoint_table}, where a Samsung Smart TV in our lab was communicating with Facebook and other advertising/tracking services.\footnote{In \Figure{endpoint_table}, the question marks on some of the domains suggest that \pioti{} did not observe DNS responses from the IP addresses (e.g., because the DNS responses were cached previously); as a result, \pioti{} had to infer the domain name based on reverse DNS or passive DNS -- hence the uncertainty.}

\begin{figure}[t] \centering
\fbox{\includegraphics[width=\columnwidth]{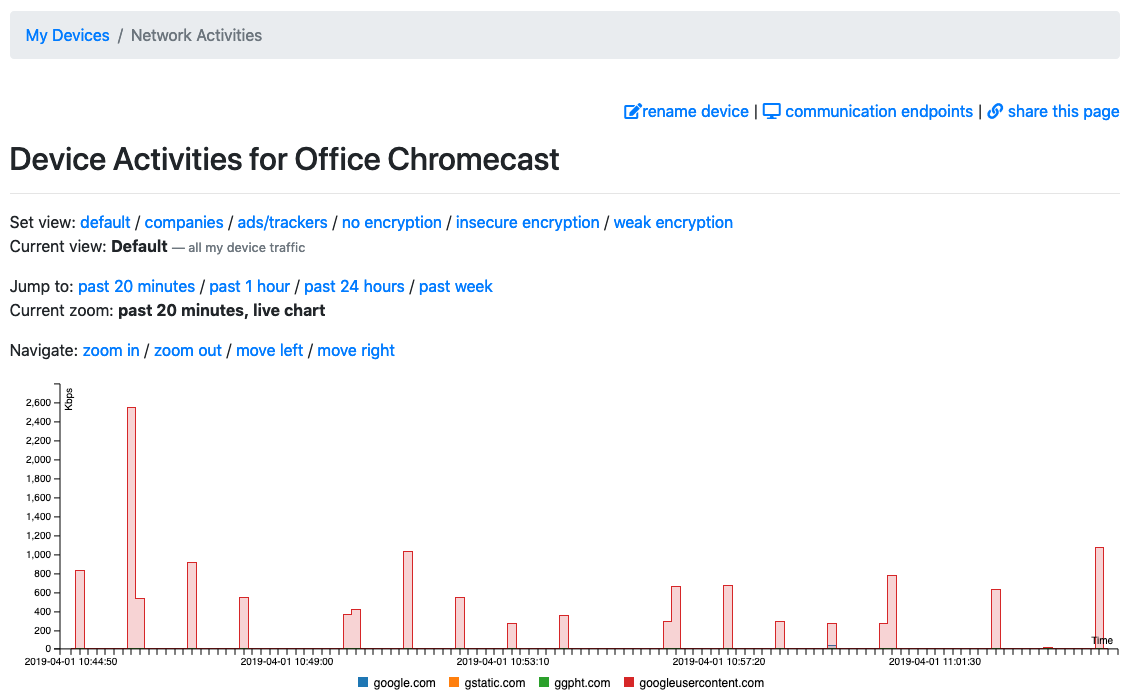}} \caption{A
screenshot of bandwidth usage for individual endpoints on a Chromecast
device.} \label{fig:chromecast_chart} \end{figure}

\paragraph{Presenting data} After the automated script labels each remote
endpoint, it generates tables and charts for the UI in real time. Each table
shows the hostnames and/or companies that a device has communicated with since
the user launched \pioti{}, e.g., \Figure{endpoint_table}. In contrast, each
chart shows the bandwidth usage by connections with individual endpoints from
a given device over time, e.g., \Figure{chromecast_chart}.

Our primary goal is for users to learn new insights about their devices,
such as what third parties a device communicates with and when devices send
and receive data. Our hope is that these insights also encourage more users to
install \pioti{} and keep running \pioti{} to monitor more devices---effectively generating a larger dataset for research.

\begin{table}[t]
\centering
\small
\begin{tabularx}{\columnwidth}{r X}
  \toprule
  \TableHeader{Category Labels} & \TableHeader{Examples}  \\
  \midrule
  appliance & Smart appliances, e.g., thermostats and vacuums \\
  tv & Smart TVs, e.g. Roku TV \\
  voice & Voice assistants, e.g., Google Home \\
  camera & Security cameras, e.g., Amazon Ring \\
  hub & IoT control hubs, e.g., Samsung SmartThings hub \\
  plug & Smart plugs/switches, e.g., Belkin Wemo switch \\
  office & Office appliances, e.g., printers and scanners \\
  storage & Network-attached storage devices (NAS) \\
  game & Game consoles, e.g., Sony PlayStation \\
  car & Internet-connected cars and peripherals \\
  computer & General computing devices, e.g., PCs and phones \\
  other & All devices not labeled above \\
  \bottomrule
\end{tabularx}
\caption{Standardized device category labels.}
\label{tab:standard_labels}
\end{table}

\section{Labeling Smart Home Devices}
\label{sec:labels}

In the previous section, we described the design and implementation of \pioti{} for the collection of network traffic data.
Before we use the dataset for research, we first standardize the user-provided labels (Section~\ref{sec:labels:standardizing}).
We then describe how we validate the correctness of the standardized labels (Section~\ref{sec:labels:validation}).


\subsection{Standardizing category and vendor labels}
\label{sec:labels:standardizing}

The user-entered category and vendor labels have two initial problems.

First, users may assign devices to fragmented categories. As shown in \Figure{device_label} and described in \Section{collecting_traffic_and_labels}, users can either select from a dropdown  the name, category, and vendor of a device, or enter an arbitrary string in a textbox with auto-compete. However, one user may categorized their smart TV as ``TV'', while another user may categorize it as ``smart TV'' or ``television'' -- apparently ignoring the auto-complete dropdown lists. As such, we manually analyze each user-entered category and standardize it as one of the labels in \Table{standard_labels} (similar to a previous study~\cite{usenix_zakir_iot}).

Second, users may have assigned devices to inconsistent categories, as users tend to have different mental models of device categories. Some users, for instance, consider  Google Homes as ``voice assistants'' while others consider them as ``smart speakers.'' In \Table{google_home_labels}, we show examples of user-entered category labels, which vary substantially across users. In light of this issue, we pick the most salient feature of the device\footnote{We use our best judgement to decide the most salient feature of a device.}
as the main category and assign one of the labels in \Table{standard_labels} -- which, for Google Home, is ``voice.'' In comparison, we label smart TVs with voice assistant features, such as Amazon Fire TV, as ``tv'' instead.

Similar problems also occur with vendor labels. For example, users entered the vendor label of Nest Cameras as both ``Nest'' and ``Google.'' We standardize the label as ``Google.''

\begin{table}[t]
\centering
\small
\begin{tabularx}{\columnwidth}{X r r}
  \toprule
  \TableHeader{User Labels} & \TableHeader{\# of Devices} & \TableHeader{\% of Google Homes} \\
  \midrule
  voice assistant & 107 & 24.8\% \\ 
smart speaker & 79 & 18.3\% \\ 
home assistant & 41 & 9.5\% \\ 
google home mini & 34 & 7.9\% \\ 
google home & 34 & 7.9\% \\ 
voice assistant speaker & 27 & 6.2\% \\ 
home mini & 13 & 3.0\% \\ 
smart speaker assistant & 11 & 2.5\% \\ 
home & 9 & 2.1\% \\ 
google mini & 7 & 1.6\% \\ 

  \bottomrule
\end{tabularx}
\caption{Top 10 user-entered category labels for Google Homes by the number of devices for each label. In total, our dataset contains 432 Google Homes.}
\label{tab:google_home_labels}
\end{table}

\subsection{Validating device labels}
\label{sec:labels:validation}

The category and vendor standardization process is based on the original name, category, and vendor labels as entered by users. Still, either or both labels can be incorrect. In this section, we describe a method to validate the standardized labels against external information, highlight the challenges of this method, and provide statistics about the distribution of devices across category and device labels.

\paragraph{Validation methods} We use six sources of information  to help us validate category and vendor labels.

\textbf{(1) FingerBank}, a proprietary API~\cite{fingerbank} that takes the OUI of a device, user agent string (if any), and five domains contacted by the device; it returns a device's likely name (e.g., ``Google Home'' or ``Generic IoT'').

\textbf{(2) Netdisco}, an open-source library that scans for smart home devices on the local network using SSDP, mDNS, and UPnP~\cite{netdisco}. The library parses any subsequent responses into JSON strings. These strings may include a device's self-advertised name (e.g., ``google\_home'').

\textbf{(3) Hostname} from DHCP Request packets. A device being monitored by \pioti{} may periodically renew its DHCP lease; the DHCP Request packet, if captured by \pioti{}, may  contain the hostname of the device (e.g., ``chromecast'').

\textbf{(4) HTTP User Agent} (UA) string. \pioti{} attempts to extract the UA from every unencrypted HTTP connection. If, for instance, the UA contains the string ``tizen,'' it is likely that the device is a Samsung Smart TV.

\textbf{(5) OUI}, extracted from the first three octets of a device's MAC address. We translate OUIs into names of manufacturers based on the IEEE OUI database~\cite{ieee-oui}. We use OUI to validate device vendor labels only and not device category labels.

\textbf{(6) Domains}: a random sample of five registered domains that a device has ever contacted, based on the traffic collected by \pioti{}. If one of the domains appears to be operated by the device's vendor (e.g., by checking the websites associated with the domains), we consider the device to be validated.



Our goal is to validate a device's standardized category and vendor labels using each of the six methods above. However, this process is difficult to fully automate. In particular, FingerBank's and Netdisco's outputs, as well as the DHCP hostnames and UAs strings, have a large number of variations; it would be a significant engineering challenge to develop regular expressions to recognize these data and validate against our labels.

Furthermore, the validation process often requires expert knowledge. For instance, if a device communicates with \texttt{xbcs.net}, we can validate the device's ``Belkin'' vendor label from our experience with Belkin products in our lab. However, doing such per-vendor manual inference at scale would be difficult.

Given these challenges, we randomly sample at most 50 devices from each category (except ``computers'' and ``others''). For every device, we manually validate the category and vendor labels using each of the six methods (except for OUI and domains, which we only use to validate vendor labels). This random audit approximates the accuracy of the standardized labels without requiring manual validation of all 8,131 labeled devices.

For each validation method, we record the outcome for each device as follows:

\begin{prettylist}

\item \textbf{No Data}. The data source for a particular validation method is missing. For instance, some devices do not respond to SSDP, mDNS, or UPnP, so Netdisco would not be applicable. In another example, a user may not have  run \pioti{} long enough to capture DHCP Request packets, so using DHCP hostnames would not be applicable.

\item \textbf{Validated}. We successfully validated the category and vendor labels using one of the six methods -- except for OUI and domains, which we only use to validate vendor labels.

\item \textbf{Not Validated}. The category and/or vendor labels are inconsistent with the validation information because, for instance, the user may have made a mistake when entering the data and/or the information from the validation methods is wrong. Unfortunately, we do not have a way to distinguish these two reasons, especially when the ground truth device identity is absent. As such, ``Not Validated'' does not necessarily mean that the user labels are wrong.

\end{prettylist}

\begin{figure}[t]
\centering
\includegraphics[width=\columnwidth]{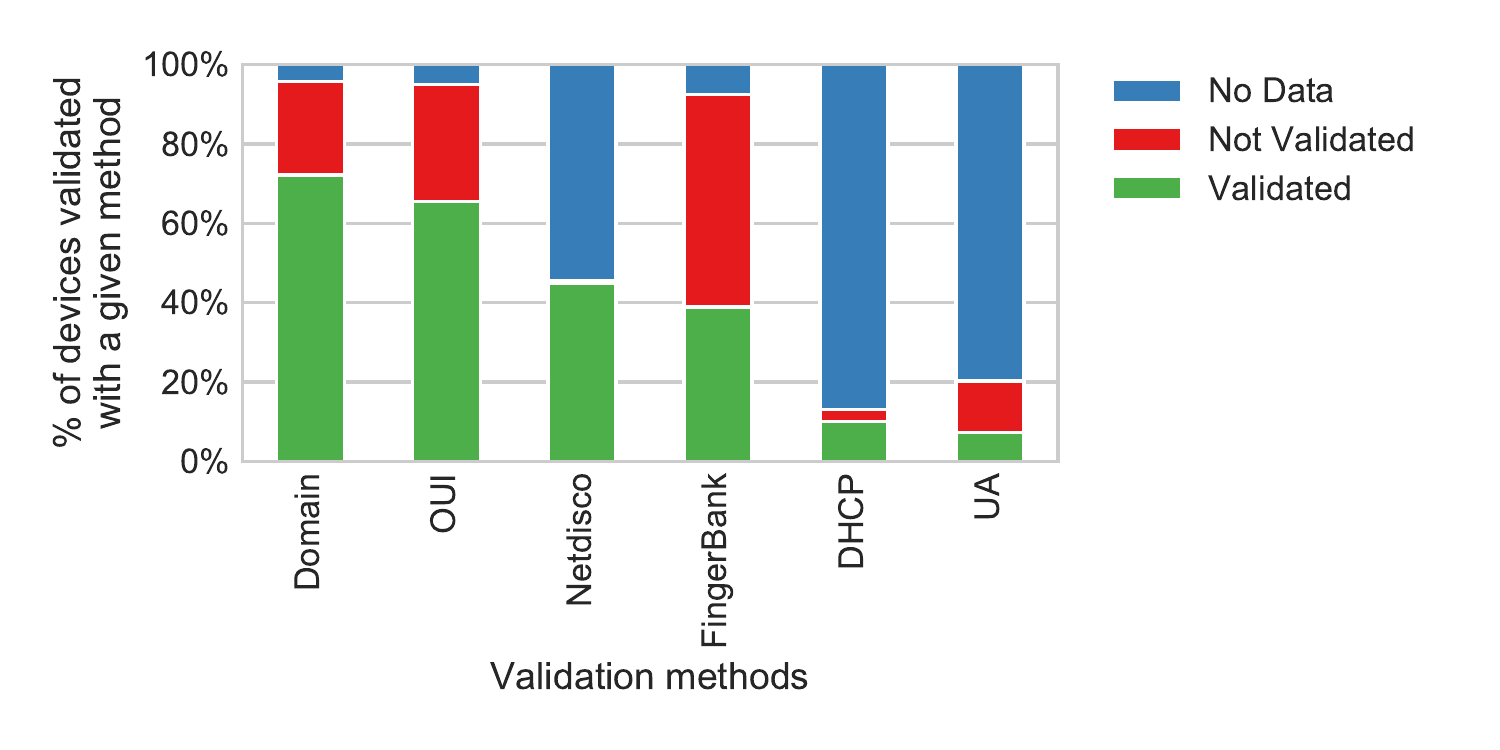}
\caption{Percentage of the 522 sampled devices whose vendor or category labels we can manually validate using each of the six validation methods. }
\label{fig:validation_over_each_method}
\end{figure}

\paragraph{Results of validation} In total, we manually sample 522 devices from our dataset: 22 devices in the ``car'' category (because there are only 22 devices in the ``car'' category) and 50 devices in each of the remaining 10 categories. \Figure{validation_over_each_method} shows the percentage of devices whose category or vendor labels we have  manually validated using each of the six validation methods.

One key takeaway from \Figure{validation_over_each_method} is that there are trade-offs  between the availability of a validation method and its effectiveness. For example, the Netdisco method is available on fewer devices than the Domain method, but Netdisco is able to validate more devices. As shown on \Figure{validation_over_each_method}, we can validate 72.2\% of the sampled devices using Domains but only 45.0\% of the sampled devices using Netdisco. One reason for this difference is that only 4.2\% of the sampled devices do not have the domain data available, whereas 54.4\% of the sampled devices did not respond to our Netdisco probes and thus lack  Netdisco data.
If we ignore devices that have neither domain nor Netdisco data, 75.4\% of the remaining devices can be validated with domains, and 98.7\% can be validated with Netdisco. These results suggest that although Netdisco data is less prevalent than domain data, Netdisco is more effective for validating device labels.

Despite their availability now, domain samples may not be the most prevalent data source for device identity validation in the near future, because domain names will likely be encrypted. In particular, DNS over HTTPS or over TLS is gaining popularity, making it difficult for an on-path observer to record the domains contacted by a device.
Moreover, the SNI field -- which includes the domain name -- in TLS Client Hello messages may be encrypted in the future~\cite{esni}.
These technological trends will likely require future device identification techniques to be less reliant on domain information.

Another observation from \Figure{validation_over_each_method} is that we cannot validate a high percentage of devices using certain methods -- e.g., 53.6\% of devices are ``Not Validated'' by FingerBank. Without any ground-truth knowledge of the device identities, we are unable to attribute this outcome to user errors or FingerBank's errors (e.g. FingerBank is unable to distinguish Google Homes and Google Chromecast during our lab tests). Given this limitation, we do not discard any devices from the dataset just because we cannot validate their labels. We defer to future work to improve device identification (\Section{future}).

\section{Dataset}
\label{sec:overview}


On April 10, 2019, we announced the release of \pioti{} with the help of social media. In particular, we first set up a website where the public would be able to download \pioti{}'s executable or source code, along with documentation on how the software works, how it collects the data, and how we use the data. We host the website on a subdomain under our academic institution (i.e., \url{https://iot-inspector.princeton.edu}). We also published Twitter posts that linked to the website. Subsequently, we observed articles about \pioti{} from a number of media outlets, including three US-based technology websites (i.e., Gizmodo, TechCrunch, and HackerNews) and the Canadian Broadcasting Corporation.

At the time of writing, \pioti{}'s dataset includes 8,488 users who have scanned 152,460 devices. Some 4,322 of these users (i.e., \textit{active users}, based on the definition in \Section{overview:users}) have allowed \pioti{} to capture network traffic from 44,956 of the devices. These numbers are still growing, as \pioti{} is actively gathering labeled traffic data.

For this paper, we analyze a subset of the data collected between April 10 and May 5, 2019. This section describes aggregate statistics about the users and devices during the 26-day study period.

\subsection{User statistics}
\label{sec:overview:users}

\paragraph{All users} Every time \pioti{} is installed, it generates and writes to disk a random User ID that persists across computer reboots. During this 26-day study period, \pioti{} collected  6,069 such unique IDs, which we assume is the number of \pioti{} users.

\begin{figure}[t]
\centering
\includegraphics[width=\columnwidth]{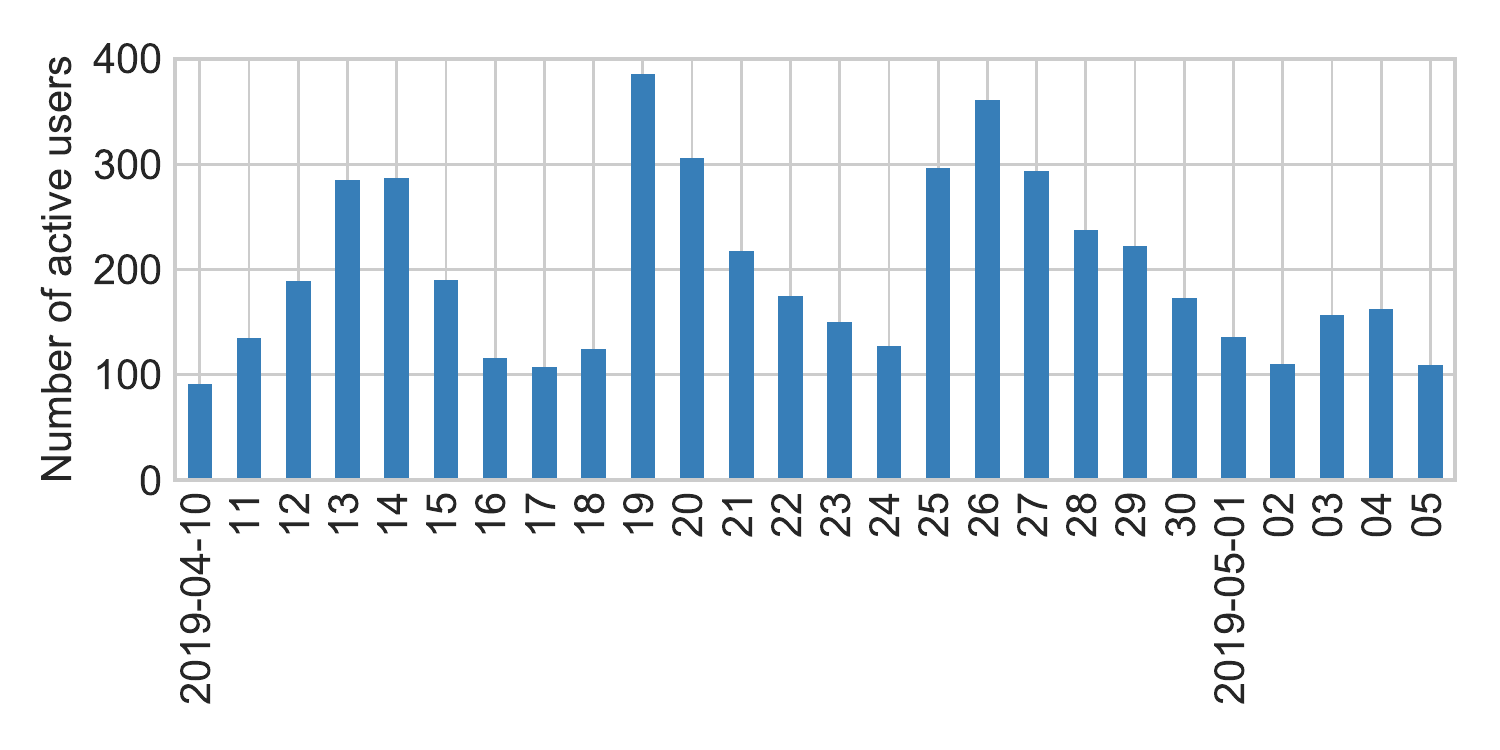}
\caption{Number of daily active users. All dates are based on UTC.}
\label{fig:daily_active_user_count}
\end{figure}

\paragraph{Active users} However, not every instance of \pioti{} was able to upload network traffic to our data collection server. In fact, only 3,388 users (55.8\% of all users) uploaded some network traffic to our server; we consider these users as \textit{active users}. We fail to observe any network traffic from the non-active users because some home routers may have dropped ARP-spoofing traffic, our data collection server may have been temporarily out of service due to high load,\footnote{\pioti{}, which runs on users' computers, stops collecting traffic if it is unable to contact the remote data collection server.} or the users may not have consented to data collection.

We observe a steady number of active users every day. During the 26-day period, there were 197.9 active daily users on average, or 174.0 users in the median case. We show the distribution of the number of active users in \Figure{daily_active_user_count}. We note that the number of active users tended to be higher around weekends (e.g., Saturdays April 13, 20, and 27).

Many of these active users ran \pioti{} for a limited duration. Half of the active users collected at least 35.3 minutes of network traffic, 25\% of the active users at least 2.8 hours of traffic, and 10\% of the active users at least 12.4 hours of traffic.

These users are likely from around the world. Even though \pioti{} does not collect users' public IP addresses, we can still infer their geographical distribution based on each user's timezone. In particular, the timezones for 64.1\% of the users are between UTC -07:00 and -04:00 (e.g., between San Francisco and New York), and for 28.0\% of the users the timezones are between UTC 00:00 and 03:00 (between London and Moscow).

\subsection{Device statistics}
\label{sec:overview:devices}

\paragraph{All devices} Upon launch, \pioti{} scans the local network and presents the user with a list of devices on the network. From April 10 to May 5, 2019, \pioti{} discovered 113,586 devices in total -- 8/15/26 devices per user in the 25th/50th/75th percentile.\footnote{For performance reasons, \pioti{} discovers at most 50 devices per user.}
For each of these devices, \pioti{} only collects the OUI and mDNS (if available) data.

\paragraph{Devices from which \pioti{} collected traffic} By default, \pioti{} does not collect network traffic from any device unless the user explicitly chooses to monitor it (\Figure{devices}). As such, \pioti{} only collected network traffic from 35,961 (31.7\%) of the discovered devices during this 26-day analysis period (i.e., across the 3,388 active users).

\paragraph{Devices with traffic and labels}
For the majority of the monitored devices, we do not have any user-entered category and vendor labels. In total, only 8,131 devices (7.2\% of all devices, 22.6\% of all monitored devices) have such labels, as entered by 1,501 users (24.7\% of all users).

For the rest of the paper, we will \textit{only examine the network traffic data from these 8,131 devices}, as their human-entered labels help us characterize security and privacy issues for particular device categories and vendors. Even with just these labeled devices, we are still looking at the largest known dataset of traffic and device labels of smart home networks in the wild.\footnote{To our knowledge, the previously largest labeled dataset of smart home traffic from lab settings consists of 50 devices over a 13-day period collected by Alrawi et al.~\cite{iot_sok}.}

\begin{table}[t]
\centering
\small
\begin{tabularx}{\columnwidth}{X r r r}
  \toprule
  \TableHeader{Cat} & \TableHeader{\# of Devices} & \TableHeader{\# of Vendors} & \TableHeader{Most Common Vendors}  \\
  \midrule
  appliance & 1,088 & 25 & google (25.3\%), ecobee (9.5\%) \\ 
tv & 984 & 19 & google (26.3\%), roku (15.2\%) \\ 
voice & 883 & 2 & amazon (51.1\%), google (48.9\%) \\ 
camera & 754 & 18 & wyze (18.2\%), amazon (16.7\%) \\ 
media & 614 & 22 & sonos (50.5\%), denon (4.6\%) \\ 
hub & 567 & 12 & philips (45.3\%), logitech (18.0\%) \\ 
plug & 553 & 12 & belkin (40.7\%), tp-link (16.6\%) \\ 
office & 185 & 5 & hp (56.8\%), epson (13.0\%) \\ 
storage & 185 & 8 & synology (51.4\%), microsoft (13.0\%) \\ 
game & 161 & 3 & nintendo (41.0\%), sony (39.8\%) \\ 
car & 22 & 3 & tesla (68.2\%), emotorwerks (9.1\%) \\ 
computer & 1,292 & 19 & apple (50.7\%), raspberry (12.0\%) \\ 
other & 843 & 34 & ubiquiti (6.5\%), eero (3.9\%) \\ 

  \bottomrule
\end{tabularx}
\caption{Overview of devices in our dataset. For each device category, we show the number of devices, number of distinct vendors, and the two vendors associated with the highest number of devices in each category.}
\label{tab:device_overview}
\end{table}

\begin{table}[t]
\centering
\small
\begin{tabularx}{\columnwidth}{X r r r}
  \toprule
  \TableHeader{Vendors} & \TableHeader{\# of Devices} & \TableHeader{\# of Vendors} & \TableHeader{Most Common Categories}  \\
  \midrule
  google & 1,066 & 5 & voice (40.5\%), appliance (25.8\%) \\ 
amazon & 733 & 7 & voice (61.5\%), tv (19.2\%) \\ 
sonos & 310 & 1 & media (100.0\%) \\ 
philips & 265 & 3 & hub (97.0\%), tv (2.3\%) \\ 
belkin & 226 & 2 & plug (99.6\%), appliance (0.4\%) \\ 
samsung & 204 & 6 & tv (58.3\%), hub (27.9\%) \\ 
roku & 152 & 2 & tv (98.7\%), media (1.3\%) \\ 
sony & 146 & 3 & game (43.8\%), tv (38.4\%) \\ 
wyze & 137 & 1 & camera (100.0\%) \\ 
xiaomi & 130 & 6 & appliance (72.3\%), camera (10.8\%) \\ 

  \bottomrule
\end{tabularx}
\caption{Vendors with the most number of devices. For each device vendor, we show the number of devices, number of distinct categories, and the two categories associated with the highest number of devices in each vendor.}
\label{tab:device_overview_vendor}
\end{table}

\paragraph{Distribution of devices across labels} For these 8,131 devices, we standardize the labels (\Section{labels:standardizing}) and count the number of devices in each category and with each vendor. Our dataset includes a diverse set of device types and vendors, as illustrated in \textbf{Tables \ref{tab:device_overview}} and \textbf{\ref{tab:device_overview_vendor}}. In total, there are 13 distinct categories and 53 unique vendors. Both tables show a diverse set of device categories and vendors in our data. Across our users, smart appliances, TVs, and voice assistants are the top three categories with the most devices. Across vendors, Google and Amazon have the most devices. Our dataset also includes vendors with a relatively small number of devices, such as eMotorWerks, which manufacturers smart chargers for electric cars, and Denon, which makes media players.

\subsection{Data release}
\label{sec:data-release}

Interested researchers can contact us to get access to the \pioti{} dataset as the following CSV files:
\begin{prettylist}
  \item \textit{Device\_labels.csv}. Columns: device identifier, category, and vendor
  \item \textit{Network\_flows.csv}. Columns: device identifier, timestamp of first packet, remote IP/hostname, remote port, protocol (i.e., TCP or UDP), number of bytes sent in a five-second window, and number of bytes received in the window.
  \item \textit{TLS\_client\_hello.csv}. Columns: device identifier, timestamp, TLS version, cipher suites, and TLS fingerprint (\Section{encryption}).
\end{prettylist}

\section{New Possibilities with Labeled Traffic Data at Scale}
\label{sec:findings}

\pioti{}'s device traffic and labels dataset -- the largest of its kind that we are aware of -- can create new research opportunities in diverse areas.
In this section, we show two examples of questions that we can answer using our data: whether smart devices encrypt their traffic using up-to-date TLS implementations; and whether they communicate with third-party advertisement services and trackers.

\subsection{Which devices encrypt their traffic?}
\label{sec:encryption}

Major browsers such as Google Chrome encourage websites to adopt HTTPS by labelling plain HTTP websites as ``insecure''~\cite{google_insecure}. However, such an effort to push for encrypted communication is yet to be seen across smart home device vendors. Our goal is to understand if TLS, a common approach to encrypting network traffic, is deployed on smart home devices and whether the encryption follows secure practices, such as using the latest TLS version and not advertising weak ciphers.

To this end, we analyze TLS ClientHello messages in the dataset. Even though the literature abounds with large TLS measurements of websites~\cite{durumeric2017security} and mobile apps~\cite{narseo-conext}, much of the existing work on smart home TLS is restricted to a small number of devices in lab settings~\cite{iot_sok}. By using the \pioti{} dataset, we provide the first and largest analysis on TLS usage for smart home devices in the wild.

\paragraphnocolon{Which devices use TLS?} We first compare the number of devices that encrypt their traffic with respect to those that do not encrypt their traffic. Specifically, we count the number of devices from which we can extract TLS Client Hello messages, which mark the beginning of a TLS connection (regardless of the port number). We compare this number with the number of devices that communicate with port 80 on the remote host, which is likely unencrypted HTTP traffic. This comparison serves as a proxy for which devices --- and also which vendors --- likely send unencrypted vs encrypted traffic.

\begin{figure}[t]
\centering
\includegraphics[width=\columnwidth]{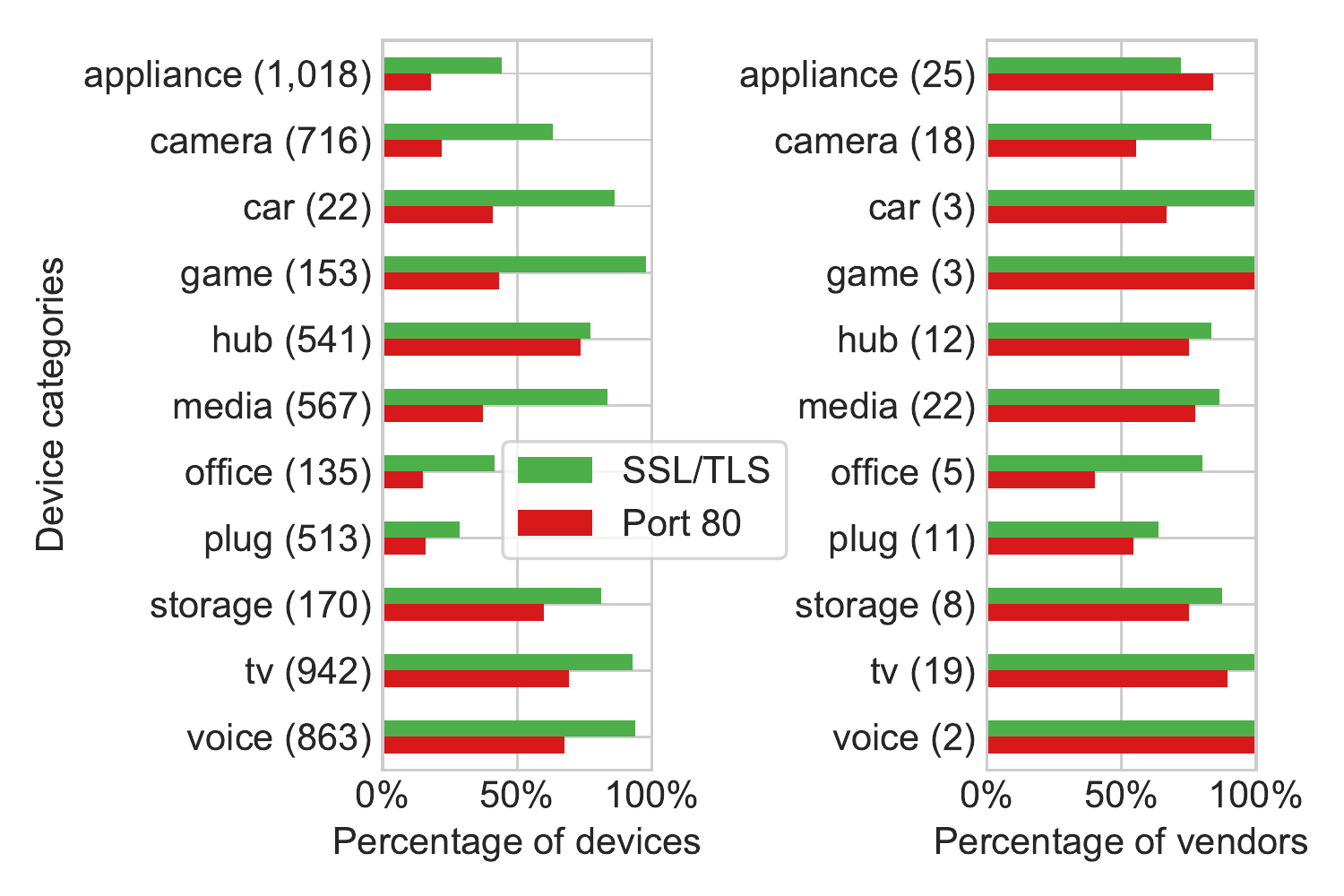}
\caption{Percentage of devices and vendors that communicate with Internet hosts via port 80 (presumably over unencrypted HTTP) or TLS. Each number in parentheses counts the number of devices (left) or vendors (right) in the respective categories. Note that a device can send either or both of encrypted and encrypted traffic.}
\label{fig:http_https}
\end{figure}

As shown in \Figure{http_https}, more devices and vendors use TLS than unencrypted HTTP. In particular, the left-hand chart shows the number of devices, thereby taking into account the purchase behaviors of our users. The right-hand chart shows the number of vendors. In total, 3,454 devices send encrypted traffic over TLS, as opposed to 2,159 devices that communicate over port 80 (presumably over unencrypted HTTP). Likewise, devices from 46 vendors use TLS, whereas devices from 44 vendors use port 80. Note that the traffic of a device can be over port 80 only, over TLS only, or both.


It is possible that we do not observe TLS traffic on certain devices. For instance, the Geeni Lightbulbs in our dataset connect with remote hosts on ports 80, 53 (DNS), and 1883 (MQTT), a lightweight messaging protocol. Despite the absence of TLS traffic, we do not know if the MQTT traffic is encrypted (e.g., over STARTTLS), because \pioti{} does not collect the payload.

On the other hand, we observe four vendors that communicate over TLS but never connect to remote port 80, including Chamberlain (which makes garage door openers), DropCam (which makes surveillance cameras), Linksys, and Oculus (which makes virtual-reality game consoles). Absent packet payload, we do not know if these devices sent unencrypted traffic on other ports.

\paragraphnocolon{Which devices use outdated TLS versions?} Even if a device communicates over TLS, the TLS implementation may not follow best practices. A smart home vendor may use an outdated or non-standard TLS library, or a vendor could (inadvertently) configure the library with insecure settings.

We focus on two examples of insecure practices that past researchers have exploited and that can potentially lead to vulnerabilities: outdated TLS versions and weak ciphers~\cite{rc4rfc,beast_attack}.

We first investigate outdated TLS versions. While the latest TLS version is 1.3, the industry-accepted version is 1.2. Versions below 1.2 are known to have vulnerabilities. For instance, TLS 1.0 is subject to the BEAST attack~\cite{beast_attack}. Although we are not aware of any actual BEAST attacks on smart home devices, we argue that any outdated TLS versions are potential targets for attacks in the future.

\begin{table}[t]
\centering
\small
\begin{tabularx}{\columnwidth}{X r | r | r r r | r r r}
  \toprule
  & \textit{\# of} & \multicolumn{1}{c|}{SSL} & \multicolumn{3}{c|}{TLS} & \multicolumn{3}{c}{Weak Ciphers} \\
  \textit{Vendor} & \textit{Devices} & \textit{3.0} & \textit{1.0} & \textit{1.1} & \textit{1.2} & \textit{Null} & \textit{RC4} & \textit{Anon.} \\
  \midrule
  vendor 1 & 669 & 0 & 31 & 0 & 645 & 0 & 9 & 15 \\
vendor 2 & 614 & 5 & 67 & 1 & 613 & 1 & 397 & 1 \\
sonos & 186 & 0 & 0 & 0 & 186 & 0 & 0 & 0 \\
vendor 4 & 148 & 1 & 41 & 0 & 129 & 0 & 80 & 1 \\
vendor 5 & 135 & 0 & 9 & 0 & 135 & 0 & 130 & 0 \\
vendor 6 & 96 & 3 & 29 & 8 & 91 & 0 & 35 & 27 \\
vendor 7 & 69 & 0 & 0 & 0 & 69 & 0 & 34 & 2 \\
wyze & 53 & 0 & 0 & 0 & 53 & 0 & 0 & 0 \\
vendor 9 & 51 & 0 & 0 & 4 & 47 & 0 & 1 & 0 \\
vendor 10 & 46 & 0 & 5 & 0 & 41 & 0 & 5 & 0 \\

  \bottomrule
\end{tabularx}
\caption{The number of devices by each vendor that use a particular TLS version and which propose a specific weak cipher in the Client Hello messages. Here, we are showing ten vendors with the most number of devices. \textit{\# of Devices} shows the number of devices that send out Client Hellos in SSL 3.0 or TLS 1.0-1.2. A device may be associated with multiple TLS versions and/or weak ciphers.}
\label{tab:ssl_tsl_weak_ciphers}
\end{table}

To understand the distribution of TLS versions across devices in our dataset, we analyze the Client Hello versions. \Table{ssl_tsl_weak_ciphers} shows ten vendors with the most observed devices that use TLS.\footnote{We are in the process of reporting the vulnerabilities to the respective vendors.} For each vendor, we count the number of devices that use a particular TLS version. Note that a device may communicate using multiple TLS versions. For instance, Vendor 2's TVs, as confirmed in our lab, communicate with Vendor 2's servers using both TLS 1.0 and 1.2. Some vendors, such as Vendor 7 (which makes network-attached storage devices) and Wyze (which makes cameras), only use TLS 1.2. In total, out of these top 10 vendors, we observe 7 vendors whose devices use TLS versions below 1.2..

\paragraphnocolon{Which devices advertise weak ciphers?} Another potentially insecure practice is advertising weak ciphers in Client Hello messages, possibly due to insecure TLS libraries or vendors' insecure settings. We look for the use of four weak ciphers, similar to those discussed in previous works~\cite{narseo-conext}: (1) null ciphers, which provide no encryption and, if used, may be vulnerable to man-in-the-middle attacks; (2) anonymous ciphers (denoted as ``Anon.'' in \Table{ssl_tsl_weak_ciphers}), which do not offer server authentication if used; (3) export-grade ciphers, which use 40-bit keys (or shorter) to comply with old US export regulations; and (4) RC ciphers, vulnerable to several known attacks and which many vendors have since stopped supporting~\cite{rc4rfc}.

We count the number of devices whose Client Hello messages advertise these weak ciphers (\Table{ssl_tsl_weak_ciphers}). No devices in our dataset advertise export-grade ciphers. RC4 is the most frequently advertised weak cipher. In particular, 397 of Vendor 2's devices advertise RC4 ciphers in Client Hellos; 283 of these devices are in the ``voice'' category, and 89 in the ``TV'' category.
Out of the top 10 vendors in our dataset, only Sonos and Wyze do not advertise weak ciphers. These are also the only two vendors that do not use outdated TLS versions.

Despite the advertisement of weak ciphers in Client Hellos, none of the devices in our dataset actually communicated over a weak cipher after the TLS handshake is complete, presumably because the server is able to negotiate the use of a secure cipher. Even so, servers can be subject to downgrade attacks, and the advertisement of weak ciphers creates potential opportunities for exploits.

\paragraph{Mitigation}
Despite many calls for smart home devices to use industry standard best-practices for data encryption, our results indicate that many vendors still use vanilla HTTP or insecure SSL/TLS versions/ciphers. This is particularly discouraging, as proper use of TLS is a low bar for smart home device vendors.

One way to mitigate this problem is through device updates. Some of the devices using deprecated versions of SSL/TLS were likely released when these versions were current. However, if devices do not support remote updates, vendors are unable to issue patches when new TLS versions are released. On the other hand, even if devices do support remote updates, the new firmware would be transmitted over a potentially vulnerable communication channel.
We echo earlier recommendations~\cite{bitag} for vendors to support smart home devices after their initial deployment and to release firmware updates to address known cryptographic vulnerabilities.

\subsection{What trackers do devices communicate with?}
\label{sec:data-aggr}

Whereas \Section{encryption} looks at the security of smart home devices, this section focuses on privacy. Our goal is to understand with what third-party services smart home devices communicate, including services that serve advertisements and track user behaviors. Although there is much existing work on such third parties on the web~\cite{englehardt2016online} and mobile devices~\cite{razaghpanah2018apps}, we are the first to study this problem in the smart home ecosystem at scale.

\paragraphnocolon{Which advertisers and trackers do smart TVs communicate with?}
The number of smart TVs, including televisions and streaming sticks, has increased over the past few years. An estimated 65.3\% of United States (US) Internet users -- nearly 182.6 million people -- will have such devices in 2018~\cite{intro1}.

Smart TVs have raised several privacy concerns, because they have access to a variety of sensitive data sources including users' viewing histories, input from built-in microphones, and user account information, which they make available to third-party developers who build applications for these devices. Many of these devices also enable behavioral advertising. For some manufacturers, such as Roku, advertising has become the primary revenue stream (as opposed to device sales~\cite{intro9,intro10,intro11}). In fact, Vizio --- a smart TV manufacturer --- was recently fined by the Federal Trade Commission (FTC) for collecting users' channel viewing histories for advertising and targeting without their consent~\cite{vizio-ftc}.

Our goal is to understand which ads/tracking services smart TVs communicate with. We show that some of these services, while common on smart TVs, are less common on the web, thus highlighting the difference in the TV/web tracking ecosystem.

To identify ads/tracking domains, we check against the Disconnect list, which Firefox uses for its private browsing mode~\cite{disconnect}. The list includes registered domains (i.e., the domain name plus the top-level domain) that are known to be trackers and advertisers. For each domain, we count the number of devices in our dataset in the ``TV'' category that communicated with the domain. We also count the number of devices in the ``computer'' category that communicated with each of the Disconnect domains. This approach allows us to compare the relative popularity of each domain across the TV and web ecosystems.

\begin{table}[t]
\centering
\small
\begin{tabularx}{\columnwidth}{X r r l}
  \toprule
  \textit{Tracking Domains} & \textit{\% of TVs} & \textit{\% of Cmpters} & \textit{Ranking in Cmpters} \\
  \midrule
  doubleclick.net & 47.1\% & 49.1\% & {\tiny $ \blacksquare  \blacksquare  \blacksquare  \blacksquare  \blacksquare  \blacksquare  \blacksquare  \blacksquare  \blacksquare  \blacksquare $} \\ 
googlesyndication.com & 22.6\% & 24.7\% & {\tiny $ \blacksquare  \blacksquare  \blacksquare  \blacksquare  \blacksquare  \blacksquare  \blacksquare  \blacksquare  \blacksquare  \blacksquare $} \\ 
crashlytics.com & 18.0\% & 48.3\% & {\tiny $ \blacksquare  \blacksquare  \blacksquare  \blacksquare  \blacksquare  \blacksquare  \blacksquare  \blacksquare  \blacksquare  \blacksquare $} \\ 
scorecardresearch.com & 14.9\% & 24.5\% & {\tiny $ \blacksquare  \blacksquare  \blacksquare  \blacksquare  \blacksquare  \blacksquare  \blacksquare  \blacksquare  \blacksquare  \blacksquare $} \\ 
sentry-cdn.com & 10.9\% & 1.2\% & {\tiny $ \blacksquare  \blacksquare  \blacksquare  \blacksquare  \blacksquare  \blacksquare  \blacksquare  \blacksquare $} \\ 
samsungads.com & 10.9\% & 0.0\% & {\tiny $ \blacksquare $} \\ 
samsungacr.com & 10.6\% & 0.0\% & {\tiny $ \blacksquare $} \\ 
google-analytics.com & 10.6\% & 37.1\% & {\tiny $ \blacksquare  \blacksquare  \blacksquare  \blacksquare  \blacksquare  \blacksquare  \blacksquare  \blacksquare  \blacksquare  \blacksquare $} \\ 
omtrdc.net & 7.1\% & 14.4\% & {\tiny $ \blacksquare  \blacksquare  \blacksquare  \blacksquare  \blacksquare  \blacksquare  \blacksquare  \blacksquare  \blacksquare  \blacksquare $} \\ 
demdex.net & 7.1\% & 18.1\% & {\tiny $ \blacksquare  \blacksquare  \blacksquare  \blacksquare  \blacksquare  \blacksquare  \blacksquare  \blacksquare  \blacksquare  \blacksquare $} \\ 
duapps.com & 6.9\% & 2.6\% & {\tiny $ \blacksquare  \blacksquare  \blacksquare  \blacksquare  \blacksquare  \blacksquare  \blacksquare  \blacksquare  \blacksquare $} \\ 
imrworldwide.com & 6.3\% & 9.7\% & {\tiny $ \blacksquare  \blacksquare  \blacksquare  \blacksquare  \blacksquare  \blacksquare  \blacksquare  \blacksquare  \blacksquare  \blacksquare $} \\ 
innovid.com & 5.1\% & 3.4\% & {\tiny $ \blacksquare  \blacksquare  \blacksquare  \blacksquare  \blacksquare  \blacksquare  \blacksquare  \blacksquare  \blacksquare $} \\ 
samsungrm.net & 4.3\% & 0.0\% & {\tiny $ \blacksquare $} \\ 
fwmrm.net & 4.3\% & 2.8\% & {\tiny $ \blacksquare  \blacksquare  \blacksquare  \blacksquare  \blacksquare  \blacksquare  \blacksquare  \blacksquare  \blacksquare $} \\ 

  \bottomrule
\end{tabularx}
\caption{The percentage of TVs and computers communicating with ads/tracking domains. We show 15 domains (out of 350) that appear on the most number of TVs which are sorted by the ``\% of TVs'' column. We also show the ranking of these domains on computers, indicated by the number of black squares. Ten squares, for instance, indicates that the ranking is in the top 10\%, while one square shows that ranking is in the bottom 10\%.}
\label{tab:top_trackers_tv_pc}
\end{table}

Out of the 984 TVs across 19 vendors, 404 devices across 14 vendors communicated with ads and trackers as labeled by Disconnect.\footnote{The remaining 5 vendors cover only 8 devices. It is possible that we did not observe any communications with trackers and advertisers because of  the small sample size.} \Table{top_trackers_tv_pc} shows the 15 ads/tracking domains that communicate with the most TVs in our dataset. We also show the percentages of TVs and computers that communicate with each of these domains. These 15 domains represent the top 4.3\% of the 350 total ad/tracking domains we have observed communicating with TVs.

We compare the ranking of ads/tracking domains across TVs and computers.
Google's DoubleClick and GoogleSyndication are the top advertisers/trackers for both TVs and computers. In contrast, several ads/tracking domains are more common on TVs than the web. For instance, \texttt{fwmrm.net} is a video advertising firm owned by Comcast. While it is ranked in the top 4.3\% of TV's ad/tracking list, its ranking, based on the number of computers who have contacted each domains, is between 10--20\% on the web.

There are also advertising and tracking domains specific to smart TVs. For instance, three Samsung domains are in the least common 10\% of observed computer ad/trackers but are prevalent for observed smart TVs. Based on the website of these domains, we speculate that Samsung TVs contact them to transmit pixel information on the smart TV screen (i.e., for automatic content recognition), gather data on the users' viewing habits, and/or to serve advertisements~\cite{samsungacr}. Other vendor-specific tracking and advertising domains include \texttt{amazon-adsystem.com}, which appears on 49.2\% of the Amazon TVs in our dataset, and \texttt{lgsmartad.com}, which appears in 38.7\% of the LG TVs in our dataset.

\paragraphnocolon{What other trackers do smart home devices communicate with?} So far, we identify advertising or tracking domains based on the Disconnect list~\cite{disconnect}, which is specifically used to block such domains on the web and on smart TVs. In the non-web and non-TV domain, however, we are not aware of any blacklists that target advertisers and trackers.

To this end, we look for third-party services that could potentially aggregate data across different types of devices. One example of such services is \textit{device control platforms}. Device control platforms coordinate device control via mobile apps, collecting device status updates and allowing users control their smart home devices through their phones. Whereas some device vendors use first-party device control platforms, e.g., Samsung Camera uses its own XMPP server on \url{xmpp.samsungsmartcam.com}, other vendors may choose to use third-party platforms, such as TuYa (an MQTT~\cite{mqtt} platform based in China) or PubNub (based in California).

These platforms may be able to observe changes in device state and infer the users' behaviors and lifestyles. For example, merely keeping track of when a user turns on/off a smart plug may reveal sensitive information on a user's life habits (e.g., when they are asleep vs awake, or whether a user is at home)~\cite{noah-huang-pets}. Although we do not have evidence whether such platforms keep and/or analyze this data, by being the first to study these platforms, we are hoping to raise the awareness of the potential privacy concerns.

To identify these platforms in our dataset, we list all domains that accept connections from devices on known device-control ports, such as MQTT (1883 and 8883) and XMPP (5222 and 5223). We also look for domains that communicate with the highest number of device categories and vendors, making sure to ignore common domains such as Google, Amazon, and NTPPool (used for time synchronization).

\begin{table}[t]
\centering
\begin{tabularx}{\columnwidth}{X r r r r}
  \toprule
  \TableHeader{Device Categories} & \TableHeader{Evrythng} & \TableHeader{PubNub} & \TableHeader{TuYa} & \TableHeader{Xively}\\
  \midrule
  appliance (25) & 1 & 2 & 3 & 1 \\ 
camera (18) & 0 & 4 & 3 & 0 \\ 
hub (12) & 0 & 3 & 0 & 2 \\ 
plug (12) & 2 & 1 & 4 & 0 \\ 
storage (8) & 1 & 0 & 0 & 0 \\ 
tv (19) & 0 & 2 & 1 & 0 \\ 
voice (2) & 0 & 0 & 2 & 0 \\ 

  \bottomrule
\end{tabularx}
\caption{Number of vendors in each category whose devices communicated with given third-party device control platforms. Each number in parentheses counts the number of vendors in the respective categories.}
\label{tab:iot_controllers}
\end{table}

We identify four device control platforms, as shown in \Table{iot_controllers}. TuYa, in particular, is used by 3 vendors in the ``appliance'' category (e.g., Chamberlain, which makes garage door openers) and 4 vendors in the ``plug'' category (e.g., Belkin and Teckin). Note that we have also observed these domains on general computing devices in our dataset, presumably contacted when smart-phone apps tried to interact with the smart home devices through the cloud.

\paragraph{Mitigation}
Users or network operators may wish to block smart home devices from communicating with certain domains for privacy reasons~\cite{bloomberg_pihole}.  Off-the-shelf tools such as Pi-hole~\cite{pi-hole}, are available to prevent Internet advertisements from all devices in a home network.

However, these tools relying on domain blocking will not be universally effective, as some devices observed in the \pioti{} dataset use hardcoded DNS resolvers. In particular, out of the 244 distinct fully-qualified hostnames contacted by all Google Home devices in our dataset, 243 of them were resolved using Google's 8.8.8.8 resolver, rather than the resolver assigned by the DHCP.  The Netflix app on smart TVs is another example. Out of the 75 fully-qualified Netflix-related hostnames (containing the strings ``netflix'' or ``nflx'') contacted by smart TVs, 65 of them are resolved using 8.8.8.8, rather than the DHCP-assigned resolvers. Vendors of these TVs include Amazon, LG, Roku, and Samsung. We have analyzed Roku and Samsung TVs in the lab and are not aware of any ways for a Roku or Samsung TV user to customize DNS resolver settings on the TV. This indicates that the DNS resolver used by the Netflix app is hard coded.
The use of hard coded DNS resolvers by smart home devices means that users and network operators would need to apply more sophisticated blocking tools to prevent devices communications with specific parties.

Another reason that domain blocking may not be effective is that functionalities of devices may be disrupted when certain advertising domains are blocked. On the web, blocking ads and trackers does not typically prevent the full webpage from being rendered (with the exception of anti-ad-blocking popups). On smart TVs, in contrast, we have shown in our lab that blocking advertising or tracking domains would prevent certain TV channels (i.e., apps for smart TVs) from loading. Furthermore, if a user is to block device control platforms such as TuYa or PubNub, the user would be unable to interact with their devices from their smart-phones if the smart-phones are not on the local network.

\section{Future Work}
\label{sec:future}

\pioti{} offers an unprecedented look into the network behavior of smart home
devices in the wild. Although certain design decisions limit the scope of data
that \pioti{} can collect, the \pioti{} dataset enables a variety of follow-up research beyond just the two examples in \Section{findings}. This section presents these limitations and opportunities for future work using the \pioti{} dataset and other smart home device analysis methods.

\subsection{Improving \pioti{}}

We describe some of \pioti{}'s limitations and discuss potential ways of improvement.

\paragraph{Promoting user engagement}
The fact that 6,069 users downloaded \pioti{} within the first 26 days of
release demonstrates widespread interest and concern about smart home security
and privacy. It also shows that many users, even those with security/privacy
concerns, trust academic institutions enough to deploy research software in
their homes. However, it is difficult to have users run \pioti{} over an extended period of time; the median duration of traffic collected from the monitored devices is only 35.3
minutes (\Section{overview}).

To improve user engagement, we plan to explore alternative UI designs. Currently,
we based the design of \pioti{} on our experience with existing work on home network measurement (\Section{related}), as well as several iterations with informal focus groups at our own university. Future work could involve a
more in-depth design exercise for \pioti{}'s interface and functionality, such as conducting qualitative user studies, or instrumenting the UI to empirically understand how existing users interact with \pioti{}.

\paragraph{Collecting more data} User privacy matters. Without limitations on
data collection, many users would be unlikely to employ research tools like
\pioti{}. We chose not to collect network traffic payloads for privacy
reasons, but this limits the extent of possible analyses. For example,
researchers and consumer advocates would like to audit whether specific
devices are collecting sensitive or other personally identifiable information
and transmitting it to third parties. Such behavior might be in violation of
privacy policies, regulation (e.g., COPPA and GDPR), or simply against the
preferences of privacy-conscious users.

We also chose not to have \pioti{} implement more active tests, such as
checking whether devices verify server certificates, because these could break
TLS connections, placing user data at risk or otherwise disrupting user
experience. However, such tests are necessary
to determine whether the devices are following security best practices.

Both of these cases represent tradeoffs between the immediate security and privacy of individual users deploying \pioti{} and legitimate research interests. We chose the current implementation of \pioti{} to prioritize the privacy of individual users given our intention to publish the resulting dataset and received approval from our university's IRB.

With these trade-offs in mind, future studies with targeted research questions or more restrictive data distribution plans could choose to collect more data than the current version of \pioti{}, provided that the users are fully informed and they express explicit consent.

\subsection{Opportunities in other research areas}

While \Section{findings} shows examples of security and privacy research, there are other research areas that could benefit from \pioti{}'s large-scale traffic and label dataset.

\paragraph{Device identification}
Before analyzing the data, we manually standardized and validated device
categories and vendors  using six validation methods (\Section{labels}).
Although this process produced a sanitized dataset that allowed us to
understand device behaviors across categories and vendors, such a practice
would unlikely scale if we had collected traffic data from more devices.

We plan to explore automatic device identification in future work. Existing literature on device identification has used a variety of machine learning methods with features from network traffic rates and packet headers \cite{meidan2017profiliot, ortiz2019devicemien, miettinen2017iot} as well as acquisitional rule-based techniques~\cite{feng2018acquisitional}. We plan to extend these studies, train machine learning models on \pioti{}'s dataset, and develop new methods that would automatically infer device identities in the wild.

\paragraph{Anomaly detection}
The ability to quickly detect misbehaving devices is an important step toward reducing  threats posed by insecure smart products. Research into anomaly and intrusion detection techniques for IoT devices~\cite{zarpelao2017survey,mirsky2018kitsune} would benefit from a large training dataset of smart home device traffic from real home networks.

Although \pioti{} includes network traffic from a diverse set of devices, the current labels only indicate device identities, rather than anomalies. There were cases where users directly emailed us about anomalous behaviors that they observed -- such as two D-Link cameras that, by default, opened port 80 on the gateway and exposed themselves to the Internet -- which we were able to reproduce independently in the lab. Beyond such anecdotes, however, we do not know whether the traffic in the \pioti{} dataset was anomalous or if any devices were compromised.

We plan to expand our user labels from simply identifying devices to identifying user activities on a particular device. We could train existing anomaly and intrusion  detectors on such activity labels, so that \pioti{} could potentially recognize traffic due to user behaviors or due to malicious activities. This insight would help us  better understand the prevalence of compromised or malfunctioning devices in the wild.

\paragraph{Health monitoring} Many smart home devices are designed to monitor human medical or psychological conditions, such as sleep monitors~\cite{sleep_monitor} or smart watches that measure the heart rate~\cite{smart_watch}. While such hardware devices could help researchers or healthcare providers learn about their participants' or patients' conditions, the cost of purchasing and deploying the hardware could potentially limit the scale of such studies.

One solution is to use the home network data to augment the data collection from existing health-monitoring devices. Apthorpe et al.~\cite{noah-huang-pets} have shown that the traffic of some smart home devices is correlated with human activities, such as whether users are home or away or whether they are asleep. Using this insight, we are exploring a new approach of health monitoring using \pioti{}, where participants would install \pioti{} on their home network, notify us of their User ID (i.e., effectively de-anonymizing their network data with their explicit consent), and label their activities (e.g., sleeping, eating, and watching TV). In this way, we could potentially build a large-scale dataset of network traffic with not only labels of device identities but also user activities. Based on the user activities, we could potentially infer the physical and mental state of users (e.g., whether the user is suffering sleep deprivation, or when the user spends a long time on their phones or TVs). Using this labeled data, we could train machine learning models to help researchers and healthcare providers monitor activities and health conditions of consented subjects at scale without dedicated hardware.

\section{Conclusion}
\label{sec:conclusion}

In response to the proliferation of smart home devices and the corresponding
lack of data enabling ubiquitous computing research in this area, we
crowdsourced a dataset of smart home network traffic and device labels from
44,956 devices across 4,322 users with \pioti{}, an open-source software
tool that we designed to enable large-scale, unobtrusive data collection
from within smart home networks. To our knowledge, this dataset is the largest
(and perhaps only) of its kind. To demonstrate the potential of this dataset
to shed new insights into smart homes, we used the data to study questions
related to smart home security and privacy.  In particular, the \pioti{}
dataset enabled us to discover the transmission of unencrypted traffic by
2,159 smart home devices across 44 vendors and to identify
insecure encryption practices; we also
identified third-party trackers and data aggregators on smart TVs and a large
variety of other smart devices. These insights are the tip of the iceberg in what
this large---and growing---dataset can offer for ubiquitous computing
research across a wide range of areas from security and privacy to human behavior.

\bibliographystyle{IEEEtran}
\bibliography{\jobname}

\end{document}